\documentclass[%
 reprint,
 amsmath,amssymb,
 aps,
]{revtex4-2}

\usepackage{graphicx}
\usepackage{siunitx}

\begin{document}

\hyphenpenalty9999

\preprint{APS/123-QED}

\title{Photon counting for axion interferometry}

\author{Haocun Yu}
\affiliation{University of Vienna, Faculty of Physics, Vienna Center for Quantum Science and Technology (VCQ),\\Boltzmanngasse 5, A-1090, Vienna, Austria. }%

\author{Ohkyung Kwon}
\affiliation{University of Chicago, Chicago, Illinois 60637, USA.}

\author{Hartmut Grote}
\affiliation{School of Physics and Astronomy
, Cardiff University, Cardiff CF24 3AA, United Kingdom.}


\author{Denis Martynov}
\affiliation{University of Birmingham, School of Physics and Astronomy, Birmingham B15 2TT, United Kingdom.}%


\begin{abstract}
Axions and axion-like particles are well-motivated dark matter candidates.
We propose an experiment that uses single photon detection interferometry to search for axions and axion-like particles in the galactic halo.
We show that photon counting with a dark rate of $6 \times 10^{-6}$\,Hz can improve the quantum sensitivity of axion interferometry by a factor of 50 compared to the quantum-enhanced heterodyne readout for 5-m long optical resonators.
The proposed experimental method has the potential to be scaled up to kilometer-long facilities, enabling the detection or setting of constraints on the axion-photon coupling coefficient of $10^{-17} - 10^{-16}$\,GeV$^{-1}$ for axion masses ranging from $0.1$ to $1$\,neV.

\end{abstract}

\maketitle

\section{Introduction}
The existence of dark matter has been overwhelmingly suggested by substantial evidence from astrophysical~\cite{Sofue_2001, Markevitch_2004, Massey_2010} and cosmological observations~\cite{Bertone_2005}. 
Among various extended theories beyond the Standard Model of particle physics, QCD axions~\cite{Peccei_1977, Preskill:1982cy, Abbott:1982af, Dine:1982ah} and pseudoscalar axion-like particles (ALPs)~\cite{Graham_2013, Ringwald_2012, Peter_2006} are widely recognized as leading candidates.

We consider ALP dark matter to behave as a coherent, classical field $a$, and interact weakly with photons through the coefficient $g_{a\gamma}$:
\begin{equation}\label{eq:L}
    \mathcal{L} \subset -\frac{1}{4}g_{a\gamma}aF\tilde{F}.
\end{equation}
This interaction induces a phase velocity difference between left- and right-handed circularly polarized light, which can be accumulated and extracted using a properly designed optical cavity and laser interferometry~\cite{Melissinos_2009, DeRocco_2018, Obata_2018}.

Several experiments were proposed in the literature to search for ALPs with interferometry ~\cite{DeRocco_2018, ACBD_2019} and the first results were recently published~\cite{oshima2023results, heinze2023results}.
In previous work, the signal light is measured by beating the signal field with a strong local oscillator field. The advantage of this technique is that the measurement can be done with a standard photodetector with a high quantum efficiency. However, the readout suffers from quantum shot noise from the local oscillator field.

In this paper, we derive the sensitivity of axion interferometers using photon-counting detectors and show their capacity to enhance detector sensitivity for axion fields with a short coherence time, compared to the periods between dark clicks of single photon detectors.
Over the past few decades, the technology for detecting single photons at near-infrared wavelengths has matured significantly~\cite{Eisaman_2011, Zwiller_2021}. The idea of employing single photon detection was also introduced in the context of microwave cavity axion searches~\cite{Lamoreaux_2013} and high-power interferometers for gravitational-wave detection~\cite{mcculler2022singlephoton}.
Recent transition-edge sensors and superconducting nanowire single photon detectors (SNSPD) feature high detection efficiency exceeding \SI{90}{\percent} and exceptionally low dark count rates of $6 \times 10^{-6}$\,Hz~\cite{Marsili_2013, Reddy_20, Verman_2021}. These detectors would extend the time intervals between detector dark clicks, surpassing the coherence time of the axion field above 0.1\,neV, and hold the potential to enhance the quantum-limited sensitivity of axion interferometers.

We hereby analyse an axion interferometer which consists of a linear optical cavity and derive its optimal parameters for heterodyne and single photon readouts. Our technique is also applicable to folded cavities. We calculate the signal-to-noise ratio for linear cavities of length from 1\,m up to 10\,km and show the advantage of single photon detection compared to heterodyne readout in axion interferometry. 
The proposed approach has the potential to be integrated into both table-top experiments and existing facilities for gravitational-wave detectors, such as GEO 600~\cite{Grote_2010} and LIGO facilities~\cite{LIGO_2009, O1instrPRL2016}.

\section{Experimental setup}
\label{sec:setup}
Figure 1 shows the proposed axion interferometer with single photon detection.
A laser beam 
in P-polarisation (red) is injected into a high-finesse Fabry-P\'erot linear cavity. To maximize the signal-to-noise ratio, the bandwidth of this main cavity is designed to match one of the axion fields. It is also beneficial to keep the cavity undercoupled ($T_2 \gg T_1$) to transmit most of the signal field to the readout port, where $T_1$ and $T_2$ are transmissivities of the input and output coupler of the cavity.
Due to the presence of the ALP field, the resonating laser field in P-polarisation undergoes partial conversion to S-polarisation (blue). The signal field around the free-spectral-range of the cavity is further amplified by the optical resonance and transmitted through the output port.

When using the method of heterodyne readout, a small fraction of the pump field is converted to S-polarisation and used as a local oscillator for the readout~\cite{Martynov_2020}.
For the proposed single photon readout, the main challenge in the experiment is isolating S-polarisation photons from the P-polarisation ones in the pump field transmitting out of the cavity. 
The transmitted pump field has a photon rate of approximately $10^{20}$\,Hz, which needs to be reduced to below $10^{-6}$\,Hz to prevent false detections by the single photon readout.

Because the pump and signal fields have orthogonal polarisations, a series of polarised optics, including polarising beam splitters and linear polarisers with high extinction performance, can be used to reduce the transmitted pump field by up to 18 orders of magnitude.
Due to imperfections in the polarisation optics, further suppression of the pump field is necessary.
Since we are searching for axions at frequencies around the free-spectral-range of the main cavity, the signal fields are separated from the pump field in frequency as well. Accordingly, a series of triangular optical cavities with non-degenerate polarisation modes can serve as both frequency and polarisation filters for final signal extraction.
These cavities can be controlled with auxiliary laser beams~\cite{Staley_2014, Izumi_12, Mullavey:12}, and they need to have a bandwidth of 1/1000 of the axion frequency and a high finesse to ensure a strong suppression of the pump field by up to factor of 8. Finally, an SNSPD with a low dark count rate, enclosed in a cryostat, is used for signal detection.

\begin{figure}[t]
\includegraphics[width=0.48\textwidth]{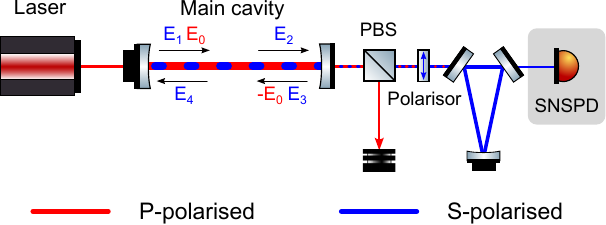}
\caption{\textbf{Schematic of the experimental setup.} P-polarised pump light resonates in a main linear optical cavity. The ALP field converts a small portion of light into S-polarised at a shifted frequency, which is isolated by a series of polarised optics and triangular optical cavities. The signal field is detected by a superconducting nanowire single photon detector (SNSPD). PBS: polarising beam splitter.}
\label{fig: setup}
\end{figure}

\section{\label{sec:level1}Sensitivity \& Integration time}

In this section, we show how single photon detectors can improve the sensitivity of axion interferometers. In our experiment, the observable quantity is the phase difference accumulated by the left- and right-handed circularly polarised light that propagates in the presence of the axion field for a time period $\tau$. In SI units, the phase difference is given by the equation
\begin{equation}\label{eq:vel}
    \Delta \phi(t, \tau) = \sqrt{\hbar c} g_{a\gamma}[a(t) - a(t-\tau)],
\end{equation}
where $\hbar$ is the Planck constant, $c$ is the speed of light, and $a(t)$ is the time-dependent amplitude of the axion field in the galactic halo.

Now, we consider how linearly polarised light propagates in the axion field between two points separated by a distance $L$. We adopt Jones calculus with the electric field vector given by $(E_p, E_s)^T$, where $E_p$ and $E_s$ are the horizontal and vertical components of the field. The Jones matrix for the propagation of light in the axion field is given by the equation
\begin{equation}
\label{eq:linear_propagation}
    A^{-1}
    \begin{pmatrix}
        e^{i\Delta \phi/2} & 0 \\
        0 & e^{-i\Delta \phi/2}
    \end{pmatrix}
     A
    \approx 
    \begin{pmatrix}
        1 & \Delta \phi/2 \\
        -\Delta \phi/2 & 1
    \end{pmatrix},
\end{equation}
where matrices $A$ and its inverse $A^{-1}$ convert electric fields from the linear basis to circular ones and back.

In further analysis, we neglect the time dependence of the pump field in the cavity $E_{\rm s, cav}$ because it is not affected by the axion field. The field in the S-polarisation builds up in the main cavity due to the axion field according to the equations
\begin{equation}
\label{eq:resonating_fields}
\begin{aligned}
        E_2(t) &= E_1(t-\tau) -
        \frac{1}{2}\Delta \phi(t, \tau) E_0, \\
        E_3(t) &= r_2 E_2(t), \\
        E_4(t) &= E_3(t-\tau) +
        \frac{1}{2}\Delta \phi(t, \tau) E_0, \\
        E_1(t) &= r_1 E_4(t),
\end{aligned}
\end{equation}
where $\tau$ is the single-trip travel time in the cavity; $r_1$ and $r_2$ are the field reflectivities of the input and output couplers; $E_0$ is the pump field forward-propagating within the main linear cavity, with its sign flipped by the output coupler; 
$E_{1-4}$ are S-polarisation electric fields propagating within the cavity near the input ($E_1$ and $E_4$) and output ($E_2$ and $E_3$) couplers (see Fig. 1). 
Solving Eq.~(\ref{eq:resonating_fields}) relative to $E_1$, we get
\begin{equation}
\label{eq:e_field}
    E_1(t) = r_1 r_2 E_1(t-2\tau) + \frac{E_0}{2}(\Delta \phi(t, \tau) - \Delta \phi(t-\tau, \tau)).
\end{equation}

If dark matter consists of ALPs with mass $m_a$, then its field behaves classically and can be written as~\cite{Budker_Casper_2014}:
\begin{equation}
\label{eq:alps_field}
    a(t) = a_0 \sin(\Omega_a t + \delta(t)),
\end{equation}
where the angular frequency $\Omega_a = 2\pi f_a = m_a c^2 / \hbar$ in SI units; $a_0 = \sqrt{2 \rho_{\rm DM}} \hbar/ m_a$ is the amplitude of the field, with $\rho_{\rm DM} \approx 5.3 \times 10^{-22}\,{\rm kg/m}^3$ as the local density of dark matter; $\delta (t)$ is the phase of the field. The phase remains constant for times $t \lesssim \tau_a$, where $\tau_a = Q_a /f_a$ is the coherence time of the field, $Q_a = c^2 / v^2 \sim 10^6$ is the quality factor of the oscillating field, and $v$ is the galactic virial velocity of the ALP dark matter~\cite{Graham_2013}. Eq.~(\ref{eq:alps_field}) neglects spacial variations of the field since ALPs wavelength $\lambda_a > 100$\,km is significantly larger than the length of the proposed experiment for $m_a < 10^{-8}$\,eV.

By setting the cavity single trip time to $\Omega_a\tau = \pi$ and applying Eq.~(\ref{eq:vel}) and (\ref{eq:alps_field}), Eq.~(\ref{eq:e_field}) can be simplified to
\begin{equation}
    E_1(t) = r_1 r_2 E_1(t-2\tau) + 2 E_0 a(t) g_{a\gamma} \sqrt{\hbar c}.
\end{equation}
Since the phase of the axion field stays constant much longer than $\tau$, the solution for the field transmitted to the readout port is given by the equation
\begin{equation}
\begin{aligned}
    E_{\rm out}(t) = \sqrt{T_2} E_1(t) &\approx E_0 \frac{\sqrt{T_2}}{1 - r_1 r_2} 2 a(t) g_{a\gamma} \sqrt{\hbar c} \\
    &\approx E_0  G \theta(t),
\end{aligned}
\end{equation}
where $G$ represents the cavity gain, $\theta \ll 1$ is the conversion efficiency of the pump field to the signal field in the presence of the axion field.

We now consider the signal-to-noise ratio accumulated with two types of readout methods: heterodyne and single photon counting. For the heterodyne readout, we intend to install a half-wave plate in the transmission path of the cavity to convert a fraction of the transmitted pump field to the local oscillator field $E_{\rm LO}$~\cite{ACBD_2019, Martynov_2020}. The time-dependent component of the power observed on a photodetector is then given by the equation
\begin{equation}
    P_h(t) = 2 E_{\rm LO} E_{\rm out}(t) = 2 \sqrt{P_{\rm LO} P_0} G \theta(t).
\end{equation}
The power spectral density of $P(t)$ around the frequency of the axion field $\Omega_a$ is given by the equation
\begin{equation}
    S_{PP} = 16 P_{\rm LO} P_0 G^2 \frac{g_{a \gamma}^2 \rho_{\rm DM} \hbar c^5}{\Omega_a^2} T,
\end{equation}
where $T = \min(T_{\rm int}, \tau_a)$ and $T_{\rm int} \gg 1/\tau$ is the integration time. For $T_{\rm int} < \tau_a$, we set our frequency spacing in the power spectral density estimation to $1/T_{\rm int}$ and the peak in $S_{PP}$ at frequency $\Omega_a$ grows linearly with time. For $T_{\rm int} \geq \tau_a$, the peak of the axion field is resolved and the power spectral density does not grow for a larger integration time. However, the signal-to-noise ratio still improves for $T_{\rm int} \geq \tau_a$ because we can subtract the mean value of the shot noise with higher precision. The power spectral density of the shot noise is given by the equation
\begin{equation}
    S_{\rm shot} = \frac{2 \hbar \omega_0 P_{\rm LO} e^{-2r}}{\sqrt{K}},
\end{equation}
where $\omega_0$ is the angular frequency of the laser light; $r$ is the squeezing factor~\cite{Caves_1985,Schnabel_2017};
$K = \max(1, T_{\rm int} / \tau_a)$ is the number of power spectral density averages that can be made with a frequency resolution of $1/T_{\rm int}$ for $T_{\rm int} < \tau_a$, or with a resolution of $1/\tau_a$ for $T_{\rm int} \geq \tau_a$.
The signal-to-noise ratio is then given by the equation
\begin{equation}
\label{eq:snr_h}
    {\rm SNR}_h^2 = \frac{S_{PP}}{S_{\rm shot}} = \frac{4 P_0 G^2 g_{a \gamma}^2 \rho_{\rm DM} c^4 \lambda e^{2r}}{\pi \Omega_a^2} \sqrt{T_{\rm int} T},
\end{equation}
where $\lambda$ is the wavelength of light.

In the case of photon counting, we observe only signal fields by rejecting the pump field in the orthogonal direction with polarisation optics and mode cleaners as discussed in Sec.~\ref{sec:setup}. The time-averaged power on the single photon detector is then given by the equation
\begin{equation}
    P_{c} = P_0 G^2 g_{a \gamma}^2 \frac{4 \rho_{\rm DM} \hbar c^5}{\Omega_a^2},
\end{equation}
and the time-averaged number of photons observed during the integration time $T_{\rm int}$ is given by the equation
\begin{equation}
    N = \frac{P_{c} T_{\rm int}}{\hbar \omega_0} = P_0 G^2 g_{a \gamma}^2 \frac{2 \rho_{\rm DM} c^4 \lambda}{\pi \Omega_a^2} T_{\rm int},
\end{equation}

We do not suffer from shot noise from the local oscillator while counting photons. However, single photon detectors observe dark counts with a time constant $\tau_d$. For state-of-the-art single photon detectors, $\tau_d$ is in the order of $10^5$\,sec. If integration time $T_{\rm int} = Q \tau_d$, we expect $Q$ dark clicks of our detector with the standard deviation of $\sqrt{Q}$. Therefore, the signal-to-noise ratio is given by the equation
\begin{equation}
\label{eq:snr_c}
    {\rm SNR}_c^2 = \frac{N}{\sqrt{Q}} = \frac{2P_0 G^2 g_{a \gamma}^2 \rho_{\rm DM} c^4 \lambda}{\pi \Omega_a^2} \sqrt{T_{\rm int} T_d},
\end{equation}
where $T_d = \min(T_{\rm int}, \tau_d)$.

Comparing the signal-to-noise ratios from Eq.~(\ref{eq:snr_h}) and Eq.~(\ref{eq:snr_c}), we find that an improvement in the estimation of $g_{a \gamma}$ can be achieved if $\tau_d \ll \tau_a$. The improvement factor is given by the equation
\begin{equation}
    \frac{{\rm SNR}_c}{{\rm SNR}_h} = \frac{1}{\sqrt{2}e^r}\left (\frac{\tau_d}{\tau_a} \right)^{1/4} \approx \frac{1}{\sqrt{2}e^r}\left(\frac{\tau_d f_a}{Q_a} \right)^{1/4}
\end{equation}

\begin{figure}[t]
\includegraphics[width=0.485\textwidth]{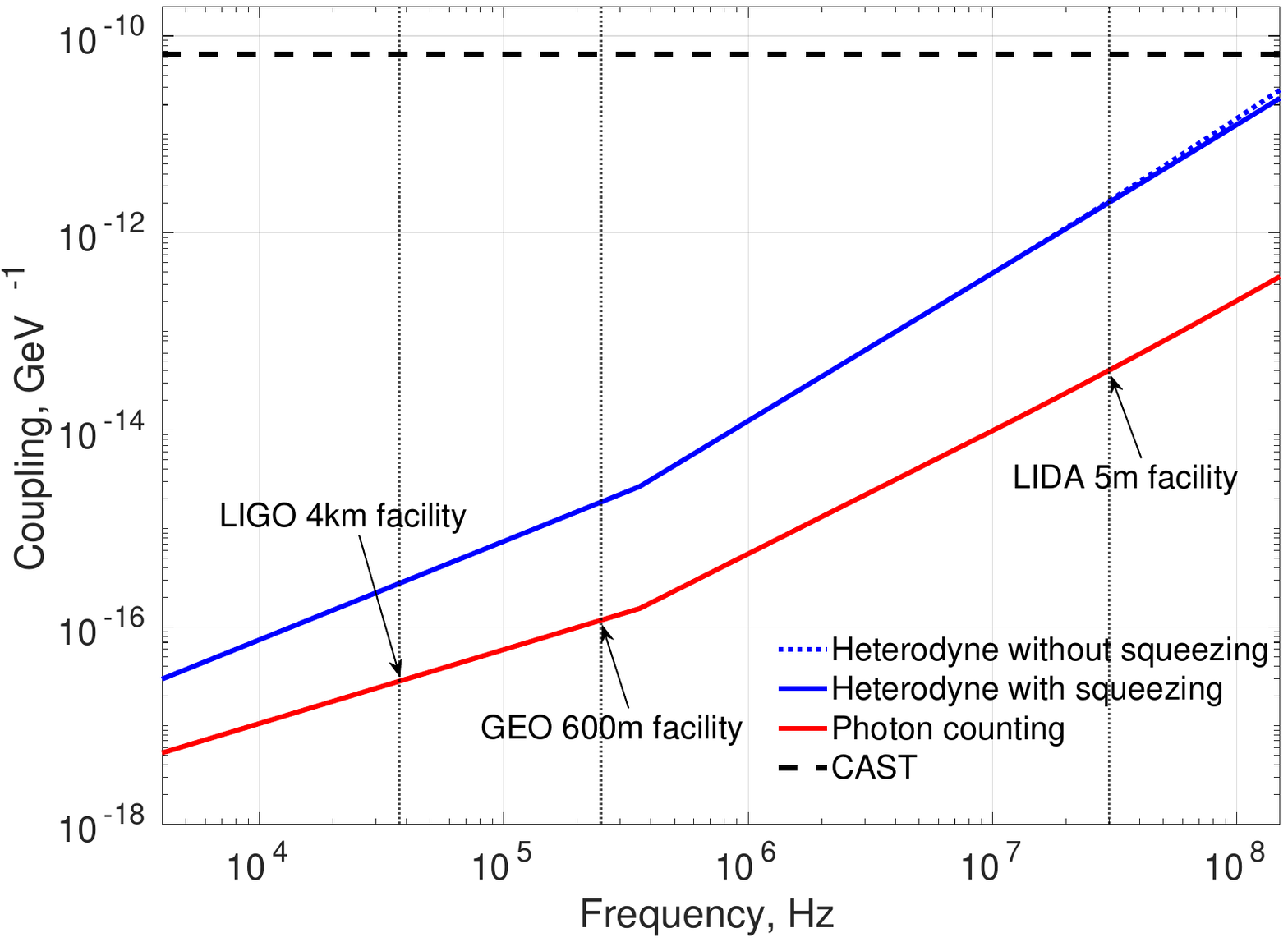}
\caption{Estimated limits on the axion-photon coupling coefficients using photon counting readout, compared with those obtained through heterodyne readout, with the CERN Axion Solar Telescope (CAST) as a reference.
}
\label{fig:sensitivity}
\end{figure}

Fig.~\ref{fig:sensitivity} compares the limits on $g_{a \gamma}$ that linear cavities can achieve with heterodyne (blue) and photon counting detectors (red). The cavity length is tuned for each frequency of the axion field $f_a = c/2L$ to satisfy the condition $\Omega_a \tau = \pi$. Based on the 5-m long LIDA detector with 120\,kW resonating power ~\cite{heinze2023results},
we assume the transmissivity of the cavity output coupler $T_2 = T_1 + Y$, where $Y = 1.7 {\, \rm ppm} \times \sqrt{L / {5 \rm m}}$ is the round-trip loss in a cavity of length $L$. 
We also impose $T_2 > 3$\,ppm to ensure that the cavity bandwidth is larger than the bandwidth of the axion field. 
We assume the resonating power of $P_0 = \min(120 {\, \rm kW} \times L / {5 \rm m}, 10\,{\rm MW})$. Since the beam size increases with the cavity length, the laser intensity on the mirrors stays the same when the cavity length and the resonating power are increased by the same factor.
The upper limit of 10\,MW is chosen to accommodate the technical complexities associated with maintaining high-quality coatings over large surface areas. 
For heterodyne readout, we assume the injection of 10\,dB squeezing when the cavity is less than 10\,m for quantum enhancement~\cite{Martynov_2020}.
For photon counting, we assume an integration time $T_{int}$ of 100\,days and $\tau_d = 1.5 \times 10^{5}$\,sec. 

Taking the 5-m LIDA detector as an example, an improvement factor of 50 can be achieved with photon counting compared to the heterodyne readout when measuring $g_{a \gamma}$ at 30\,MHz ($m_a = 124$\,neV). The length of the GEO 600 facility corresponds to an axion frequency of 250\,kHz ($m_a = 1$\,neV), resulting in an improvement factor of 16. For a 4-km detector that can be installed in the LIGO facilities, the improvement factor would be 10 for an axion frequency of 37.5\,kHz ($m_a = 0.15$\,neV).

We also note that the sensitivity curves in Fig.~\ref{fig:sensitivity} are calculated for a light wavelength $\lambda=1064$\,nm. Since Eq.~(\ref{eq:snr_c}) shows that the scaling of the signal-to-noise ratio scales as ${\rm SNR} \sim \sqrt{\lambda}$ then the reduction of wavelength leads to an improvement on the $g_{a \gamma}$ constrains. Particularly, a 5-m long GHz resonator, similar to the ADMX one, and a qubit (two-level system) operating as a single microwave photon detector~\cite{Dixit_2021} has the potential to probe axion fields around 100\,neV down to $g_{a \gamma} \approx 10^{-16}$\,GeV$^{-1}$.

\section{\label{sec:level1}Conclusion}
In this work, we proposed an axion interferometer with single photon counting methods targeting to detect or set constraints for axion-photon coupling coefficient for axion masses of $0.1 - 100$\,neV.
Looking into the future, current gravitational-wave facilities are potential infrastructures to be transformed into axion interferometers, given their existing high-power lasers, ultra-stable linear optical cavities, and vacuum envelopes.

The key challenge of separating the pump field from the signal field can be approached by installing polarisation optics and a set of mode cleaners on the readout path. We also note that the sensitivity of the axion interferometer with photon readout is limited by the dark rate of state-of-the-art single photon detectors. Anticipated advancements in single photon detector technologies will lead to enhanced constraints on the parameter $g_{a \gamma}$.

Finally, we computed the sensitivity curve for both heterodyne and single-photon readout across resonator lengths ranging from 1\,m to 10\,km. The scaling of the SNR improvement will be similar for folded resonators as long as the dark count rate is larger than the bandwidth of the axion field. The scaling is also applicable for GHz resonators, which have the potential to probe the axion-photon interaction at a deeper level than constraining $g_{a \gamma}$ with optical resonators.

\begin{acknowledgments}
We wish to acknowledge the support of the Quantum Interferometry collaboration for useful discussions.
H.Yu acknowledges support from the Marie-Skłodowska
Curie Postdoctoral Fellowship program, hosted by the
Horizon Europe. 
D.M. acknowledges the support of the Institute for Gravitational Wave Astronomy at the University of Birmingham and STFC Quantum Technology for Fundamental Physics scheme (Grant No. ST/T006609/1 and ST/W006375/1). D.M. is supported by the 2021 Philip Leverhulme Prize. 
\end{acknowledgments}

\bibliography{biblio}

\begin{thebibliography}{35}%
\makeatletter
\providecommand \@ifxundefined [1]{%
 \@ifx{#1\undefined}
}%
\providecommand \@ifnum [1]{%
 \ifnum #1\expandafter \@firstoftwo
 \else \expandafter \@secondoftwo
 \fi
}%
\providecommand \@ifx [1]{%
 \ifx #1\expandafter \@firstoftwo
 \else \expandafter \@secondoftwo
 \fi
}%
\providecommand \natexlab [1]{#1}%
\providecommand \enquote  [1]{``#1''}%
\providecommand \bibnamefont  [1]{#1}%
\providecommand \bibfnamefont [1]{#1}%
\providecommand \citenamefont [1]{#1}%
\providecommand \href@noop [0]{\@secondoftwo}%
\providecommand \href [0]{\begingroup \@sanitize@url \@href}%
\providecommand \@href[1]{\@@startlink{#1}\@@href}%
\providecommand \@@href[1]{\endgroup#1\@@endlink}%
\providecommand \@sanitize@url [0]{\catcode `\\12\catcode `\$12\catcode
  `\&12\catcode `\#12\catcode `\^12\catcode `\_12\catcode `\%12\relax}%
\providecommand \@@startlink[1]{}%
\providecommand \@@endlink[0]{}%
\providecommand \url  [0]{\begingroup\@sanitize@url \@url }%
\providecommand \@url [1]{\endgroup\@href {#1}{\urlprefix }}%
\providecommand \urlprefix  [0]{URL }%
\providecommand \Eprint [0]{\href }%
\providecommand \doibase [0]{https://doi.org/}%
\providecommand \selectlanguage [0]{\@gobble}%
\providecommand \bibinfo  [0]{\@secondoftwo}%
\providecommand \bibfield  [0]{\@secondoftwo}%
\providecommand \translation [1]{[#1]}%
\providecommand \BibitemOpen [0]{}%
\providecommand \bibitemStop [0]{}%
\providecommand \bibitemNoStop [0]{.\EOS\space}%
\providecommand \EOS [0]{\spacefactor3000\relax}%
\providecommand \BibitemShut  [1]{\csname bibitem#1\endcsname}%
\let\auto@bib@innerbib\@empty
\bibitem [{\citenamefont {Sofue}\ and\ \citenamefont
  {Rubin}(2001)}]{Sofue_2001}%
  \BibitemOpen
  \bibfield  {author} {\bibinfo {author} {\bibfnamefont {Y.}~\bibnamefont
  {Sofue}}\ and\ \bibinfo {author} {\bibfnamefont {V.}~\bibnamefont {Rubin}},\
  }\bibfield  {title} {\bibinfo {title} {Rotation curves of spiral galaxies},\
  }\href {https://doi.org/10.1146/annurev.astro.39.1.137} {\bibfield  {journal}
  {\bibinfo  {journal} {Annual Review of Astronomy and Astrophysics}\ }\textbf
  {\bibinfo {volume} {39}},\ \bibinfo {pages} {137} (\bibinfo {year}
  {2001})}\BibitemShut {NoStop}%
\bibitem [{\citenamefont {Markevitch}\ \emph {et~al.}(2004)\citenamefont
  {Markevitch}, \citenamefont {Gonzalez}, \citenamefont {Clowe}, \citenamefont
  {Vikhlinin}, \citenamefont {Forman}, \citenamefont {Jones}, \citenamefont
  {Murray},\ and\ \citenamefont {Tucker}}]{Markevitch_2004}%
  \BibitemOpen
  \bibfield  {author} {\bibinfo {author} {\bibfnamefont {M.}~\bibnamefont
  {Markevitch}}, \bibinfo {author} {\bibfnamefont {A.~H.}\ \bibnamefont
  {Gonzalez}}, \bibinfo {author} {\bibfnamefont {D.}~\bibnamefont {Clowe}},
  \bibinfo {author} {\bibfnamefont {A.}~\bibnamefont {Vikhlinin}}, \bibinfo
  {author} {\bibfnamefont {W.}~\bibnamefont {Forman}}, \bibinfo {author}
  {\bibfnamefont {C.}~\bibnamefont {Jones}}, \bibinfo {author} {\bibfnamefont
  {S.}~\bibnamefont {Murray}},\ and\ \bibinfo {author} {\bibfnamefont
  {W.}~\bibnamefont {Tucker}},\ }\bibfield  {title} {\bibinfo {title} {Direct
  constraints on the dark matter self-interaction cross section from the
  merging galaxy cluster 1e 0657-56},\ }\href {https://doi.org/10.1086/383178}
  {\bibfield  {journal} {\bibinfo  {journal} {The Astrophysical Journal}\
  }\textbf {\bibinfo {volume} {606}},\ \bibinfo {pages} {819} (\bibinfo {year}
  {2004})}\BibitemShut {NoStop}%
\bibitem [{\citenamefont {Massey}\ \emph {et~al.}(2010)\citenamefont {Massey},
  \citenamefont {Kitching},\ and\ \citenamefont {Richard}}]{Massey_2010}%
  \BibitemOpen
  \bibfield  {author} {\bibinfo {author} {\bibfnamefont {R.}~\bibnamefont
  {Massey}}, \bibinfo {author} {\bibfnamefont {T.}~\bibnamefont {Kitching}},\
  and\ \bibinfo {author} {\bibfnamefont {J.}~\bibnamefont {Richard}},\
  }\bibfield  {title} {\bibinfo {title} {The dark matter of gravitational
  lensing},\ }\href {https://doi.org/10.1088/0034-4885/73/8/086901} {\bibfield
  {journal} {\bibinfo  {journal} {Reports on Progress in Physics}\ }\textbf
  {\bibinfo {volume} {73}},\ \bibinfo {pages} {086901} (\bibinfo {year}
  {2010})}\BibitemShut {NoStop}%
\bibitem [{\citenamefont {Bertone}\ \emph {et~al.}(2005)\citenamefont
  {Bertone}, \citenamefont {Hooper},\ and\ \citenamefont
  {Silk}}]{Bertone_2005}%
  \BibitemOpen
  \bibfield  {author} {\bibinfo {author} {\bibfnamefont {G.}~\bibnamefont
  {Bertone}}, \bibinfo {author} {\bibfnamefont {D.}~\bibnamefont {Hooper}},\
  and\ \bibinfo {author} {\bibfnamefont {J.}~\bibnamefont {Silk}},\ }\bibfield
  {title} {\bibinfo {title} {Particle dark matter: evidence, candidates and
  constraints},\ }\href {https://doi.org/10.1016/j.physrep.2004.08.031}
  {\bibfield  {journal} {\bibinfo  {journal} {Physics Reports}\ }\textbf
  {\bibinfo {volume} {405}},\ \bibinfo {pages} {279} (\bibinfo {year}
  {2005})}\BibitemShut {NoStop}%
\bibitem [{\citenamefont {Peccei}\ and\ \citenamefont
  {Quinn}(1977)}]{Peccei_1977}%
  \BibitemOpen
  \bibfield  {author} {\bibinfo {author} {\bibfnamefont {R.~D.}\ \bibnamefont
  {Peccei}}\ and\ \bibinfo {author} {\bibfnamefont {H.~R.}\ \bibnamefont
  {Quinn}},\ }\bibfield  {title} {\bibinfo {title} {$\mathrm{CP}$ conservation
  in the presence of pseudoparticles},\ }\href
  {https://doi.org/10.1103/PhysRevLett.38.1440} {\bibfield  {journal} {\bibinfo
   {journal} {Phys. Rev. Lett.}\ }\textbf {\bibinfo {volume} {38}},\ \bibinfo
  {pages} {1440} (\bibinfo {year} {1977})}\BibitemShut {NoStop}%
\bibitem [{\citenamefont {Preskill}\ \emph {et~al.}(1983)\citenamefont
  {Preskill}, \citenamefont {Wise},\ and\ \citenamefont
  {Wilczek}}]{Preskill:1982cy}%
  \BibitemOpen
  \bibfield  {author} {\bibinfo {author} {\bibfnamefont {J.}~\bibnamefont
  {Preskill}}, \bibinfo {author} {\bibfnamefont {M.~B.}\ \bibnamefont {Wise}},\
  and\ \bibinfo {author} {\bibfnamefont {F.}~\bibnamefont {Wilczek}},\
  }\bibfield  {title} {\bibinfo {title} {{Cosmology of the Invisible Axion}},\
  }\href {https://doi.org/10.1016/0370-2693(83)90637-8} {\bibfield  {journal}
  {\bibinfo  {journal} {Phys. Lett. B}\ }\textbf {\bibinfo {volume} {120}},\
  \bibinfo {pages} {127} (\bibinfo {year} {1983})}\BibitemShut {NoStop}%
\bibitem [{\citenamefont {Abbott}\ and\ \citenamefont
  {Sikivie}(1983)}]{Abbott:1982af}%
  \BibitemOpen
  \bibfield  {author} {\bibinfo {author} {\bibfnamefont {L.~F.}\ \bibnamefont
  {Abbott}}\ and\ \bibinfo {author} {\bibfnamefont {P.}~\bibnamefont
  {Sikivie}},\ }\bibfield  {title} {\bibinfo {title} {{A Cosmological Bound on
  the Invisible Axion}},\ }\href {https://doi.org/10.1016/0370-2693(83)90638-X}
  {\bibfield  {journal} {\bibinfo  {journal} {Phys. Lett. B}\ }\textbf
  {\bibinfo {volume} {120}},\ \bibinfo {pages} {133} (\bibinfo {year}
  {1983})}\BibitemShut {NoStop}%
\bibitem [{\citenamefont {Dine}\ and\ \citenamefont
  {Fischler}(1983)}]{Dine:1982ah}%
  \BibitemOpen
  \bibfield  {author} {\bibinfo {author} {\bibfnamefont {M.}~\bibnamefont
  {Dine}}\ and\ \bibinfo {author} {\bibfnamefont {W.}~\bibnamefont
  {Fischler}},\ }\bibfield  {title} {\bibinfo {title} {{The Not So Harmless
  Axion}},\ }\href {https://doi.org/10.1016/0370-2693(83)90639-1} {\bibfield
  {journal} {\bibinfo  {journal} {Phys. Lett. B}\ }\textbf {\bibinfo {volume}
  {120}},\ \bibinfo {pages} {137} (\bibinfo {year} {1983})}\BibitemShut
  {NoStop}%
\bibitem [{\citenamefont {Graham}\ and\ \citenamefont
  {Rajendran}(2013)}]{Graham_2013}%
  \BibitemOpen
  \bibfield  {author} {\bibinfo {author} {\bibfnamefont {P.~W.}\ \bibnamefont
  {Graham}}\ and\ \bibinfo {author} {\bibfnamefont {S.}~\bibnamefont
  {Rajendran}},\ }\bibfield  {title} {\bibinfo {title} {New observables for
  direct detection of axion dark matter},\ }\bibfield  {journal} {\bibinfo
  {journal} {Physical Review D}\ }\textbf {\bibinfo {volume} {88}},\ \href
  {https://doi.org/10.1103/physrevd.88.035023} {10.1103/physrevd.88.035023}
  (\bibinfo {year} {2013})\BibitemShut {NoStop}%
\bibitem [{\citenamefont {Ringwald}(2012)}]{Ringwald_2012}%
  \BibitemOpen
  \bibfield  {author} {\bibinfo {author} {\bibfnamefont {A.}~\bibnamefont
  {Ringwald}},\ }\bibfield  {title} {\bibinfo {title} {Exploring the role of
  axions and other wisps in the dark universe},\ }\href
  {https://doi.org/https://doi.org/10.1016/j.dark.2012.10.008} {\bibfield
  {journal} {\bibinfo  {journal} {Physics of the Dark Universe}\ }\textbf
  {\bibinfo {volume} {1}},\ \bibinfo {pages} {116} (\bibinfo {year} {2012})},\
  \bibinfo {note} {next Decade in Dark Matter and Dark Energy}\BibitemShut
  {NoStop}%
\bibitem [{\citenamefont {Svrcek}\ and\ \citenamefont
  {Witten}(2006)}]{Peter_2006}%
  \BibitemOpen
  \bibfield  {author} {\bibinfo {author} {\bibfnamefont {P.}~\bibnamefont
  {Svrcek}}\ and\ \bibinfo {author} {\bibfnamefont {E.}~\bibnamefont
  {Witten}},\ }\bibfield  {title} {\bibinfo {title} {Axions in string theory},\
  }\href {https://doi.org/10.1088/1126-6708/2006/06/051} {\bibfield  {journal}
  {\bibinfo  {journal} {Journal of High Energy Physics}\ }\textbf {\bibinfo
  {volume} {2006}},\ \bibinfo {pages} {051} (\bibinfo {year}
  {2006})}\BibitemShut {NoStop}%
\bibitem [{\citenamefont {Melissinos}(2009)}]{Melissinos_2009}%
  \BibitemOpen
  \bibfield  {author} {\bibinfo {author} {\bibfnamefont {A.~C.}\ \bibnamefont
  {Melissinos}},\ }\bibfield  {title} {\bibinfo {title} {Proposal for a search
  for cosmic axions using an optical cavity},\ }\href
  {https://doi.org/10.1103/PhysRevLett.102.202001} {\bibfield  {journal}
  {\bibinfo  {journal} {Phys. Rev. Lett.}\ }\textbf {\bibinfo {volume} {102}},\
  \bibinfo {pages} {202001} (\bibinfo {year} {2009})}\BibitemShut {NoStop}%
\bibitem [{\citenamefont {DeRocco}\ and\ \citenamefont
  {Hook}(2018)}]{DeRocco_2018}%
  \BibitemOpen
  \bibfield  {author} {\bibinfo {author} {\bibfnamefont {W.}~\bibnamefont
  {DeRocco}}\ and\ \bibinfo {author} {\bibfnamefont {A.}~\bibnamefont {Hook}},\
  }\bibfield  {title} {\bibinfo {title} {Axion interferometry},\ }\bibfield
  {journal} {\bibinfo  {journal} {Physical Review D}\ }\textbf {\bibinfo
  {volume} {98}},\ \href {https://doi.org/10.1103/physrevd.98.035021}
  {10.1103/physrevd.98.035021} (\bibinfo {year} {2018})\BibitemShut {NoStop}%
\bibitem [{\citenamefont {Obata}\ \emph {et~al.}(2018)\citenamefont {Obata},
  \citenamefont {Fujita},\ and\ \citenamefont {Michimura}}]{Obata_2018}%
  \BibitemOpen
  \bibfield  {author} {\bibinfo {author} {\bibfnamefont {I.}~\bibnamefont
  {Obata}}, \bibinfo {author} {\bibfnamefont {T.}~\bibnamefont {Fujita}},\ and\
  \bibinfo {author} {\bibfnamefont {Y.}~\bibnamefont {Michimura}},\ }\bibfield
  {title} {\bibinfo {title} {Optical ring cavity search for axion dark
  matter},\ }\bibfield  {journal} {\bibinfo  {journal} {Physical Review
  Letters}\ }\textbf {\bibinfo {volume} {121}},\ \href
  {https://doi.org/10.1103/physrevlett.121.161301}
  {10.1103/physrevlett.121.161301} (\bibinfo {year} {2018})\BibitemShut
  {NoStop}%
\bibitem [{\citenamefont {Liu}\ \emph {et~al.}(2019)\citenamefont {Liu},
  \citenamefont {Elwood}, \citenamefont {Evans},\ and\ \citenamefont
  {Thaler}}]{ACBD_2019}%
  \BibitemOpen
  \bibfield  {author} {\bibinfo {author} {\bibfnamefont {H.}~\bibnamefont
  {Liu}}, \bibinfo {author} {\bibfnamefont {B.~D.}\ \bibnamefont {Elwood}},
  \bibinfo {author} {\bibfnamefont {M.}~\bibnamefont {Evans}},\ and\ \bibinfo
  {author} {\bibfnamefont {J.}~\bibnamefont {Thaler}},\ }\bibfield  {title}
  {\bibinfo {title} {Searching for axion dark matter with birefringent
  cavities},\ }\href {https://doi.org/10.1103/PhysRevD.100.023548} {\bibfield
  {journal} {\bibinfo  {journal} {Phys. Rev. D}\ }\textbf {\bibinfo {volume}
  {100}},\ \bibinfo {pages} {023548} (\bibinfo {year} {2019})}\BibitemShut
  {NoStop}%
\bibitem [{\citenamefont {Oshima}\ \emph {et~al.}(2023)\citenamefont {Oshima},
  \citenamefont {Fujimoto}, \citenamefont {Kume}, \citenamefont {Morisaki},
  \citenamefont {Nagano}, \citenamefont {Fujita}, \citenamefont {Obata},
  \citenamefont {Nishizawa}, \citenamefont {Michimura},\ and\ \citenamefont
  {Ando}}]{oshima2023results}%
  \BibitemOpen
  \bibfield  {author} {\bibinfo {author} {\bibfnamefont {Y.}~\bibnamefont
  {Oshima}}, \bibinfo {author} {\bibfnamefont {H.}~\bibnamefont {Fujimoto}},
  \bibinfo {author} {\bibfnamefont {J.}~\bibnamefont {Kume}}, \bibinfo {author}
  {\bibfnamefont {S.}~\bibnamefont {Morisaki}}, \bibinfo {author}
  {\bibfnamefont {K.}~\bibnamefont {Nagano}}, \bibinfo {author} {\bibfnamefont
  {T.}~\bibnamefont {Fujita}}, \bibinfo {author} {\bibfnamefont
  {I.}~\bibnamefont {Obata}}, \bibinfo {author} {\bibfnamefont
  {A.}~\bibnamefont {Nishizawa}}, \bibinfo {author} {\bibfnamefont
  {Y.}~\bibnamefont {Michimura}},\ and\ \bibinfo {author} {\bibfnamefont
  {M.}~\bibnamefont {Ando}},\ }\href@noop {} {\bibinfo {title} {First results
  of axion dark matter search with dance}} (\bibinfo {year} {2023}),\ \Eprint
  {https://arxiv.org/abs/2303.03594} {arXiv:2303.03594 [hep-ex]} \BibitemShut
  {NoStop}%
\bibitem [{\citenamefont {Heinze}\ \emph {et~al.}(2023)\citenamefont {Heinze},
  \citenamefont {Gill}, \citenamefont {Dmitriev}, \citenamefont {Smetana},
  \citenamefont {Yan}, \citenamefont {Boyer}, \citenamefont {Martynov},\ and\
  \citenamefont {Evans}}]{heinze2023results}%
  \BibitemOpen
  \bibfield  {author} {\bibinfo {author} {\bibfnamefont {J.}~\bibnamefont
  {Heinze}}, \bibinfo {author} {\bibfnamefont {A.}~\bibnamefont {Gill}},
  \bibinfo {author} {\bibfnamefont {A.}~\bibnamefont {Dmitriev}}, \bibinfo
  {author} {\bibfnamefont {J.}~\bibnamefont {Smetana}}, \bibinfo {author}
  {\bibfnamefont {T.}~\bibnamefont {Yan}}, \bibinfo {author} {\bibfnamefont
  {V.}~\bibnamefont {Boyer}}, \bibinfo {author} {\bibfnamefont
  {D.}~\bibnamefont {Martynov}},\ and\ \bibinfo {author} {\bibfnamefont
  {M.}~\bibnamefont {Evans}},\ }\href@noop {} {\bibinfo {title} {{First results
  of the Laser-Interferometric Detector for Axions (LIDA)}}} (\bibinfo {year}
  {2023}),\ \Eprint {https://arxiv.org/abs/2307.01365} {arXiv:2307.01365
  [astro-ph.CO]} \BibitemShut {NoStop}%
\bibitem [{\citenamefont {Eisaman}\ \emph {et~al.}(2011)\citenamefont
  {Eisaman}, \citenamefont {Fan}, \citenamefont {Migdall},\ and\ \citenamefont
  {Polyakov}}]{Eisaman_2011}%
  \BibitemOpen
  \bibfield  {author} {\bibinfo {author} {\bibfnamefont {M.}~\bibnamefont
  {Eisaman}}, \bibinfo {author} {\bibfnamefont {J.}~\bibnamefont {Fan}},
  \bibinfo {author} {\bibfnamefont {A.}~\bibnamefont {Migdall}},\ and\ \bibinfo
  {author} {\bibfnamefont {S.}~\bibnamefont {Polyakov}},\ }\bibfield  {title}
  {\bibinfo {title} {Invited review article: Single-photon sources and
  detectors},\ }\href {https://doi.org/10.1063/1.3610677} {\bibfield  {journal}
  {\bibinfo  {journal} {The Review of scientific instruments}\ }\textbf
  {\bibinfo {volume} {82}},\ \bibinfo {pages} {071101} (\bibinfo {year}
  {2011})}\BibitemShut {NoStop}%
\bibitem [{\citenamefont {Esmaeil~Zadeh}\ \emph {et~al.}(2021)\citenamefont
  {Esmaeil~Zadeh}, \citenamefont {Chang}, \citenamefont {Los}, \citenamefont
  {Gyger}, \citenamefont {Elshaari}, \citenamefont {Steinhauer}, \citenamefont
  {Dorenbos},\ and\ \citenamefont {Zwiller}}]{Zwiller_2021}%
  \BibitemOpen
  \bibfield  {author} {\bibinfo {author} {\bibfnamefont {I.}~\bibnamefont
  {Esmaeil~Zadeh}}, \bibinfo {author} {\bibfnamefont {J.}~\bibnamefont
  {Chang}}, \bibinfo {author} {\bibfnamefont {J.}~\bibnamefont {Los}}, \bibinfo
  {author} {\bibfnamefont {S.}~\bibnamefont {Gyger}}, \bibinfo {author}
  {\bibfnamefont {A.~W.}\ \bibnamefont {Elshaari}}, \bibinfo {author}
  {\bibfnamefont {S.}~\bibnamefont {Steinhauer}}, \bibinfo {author}
  {\bibfnamefont {S.}~\bibnamefont {Dorenbos}},\ and\ \bibinfo {author}
  {\bibfnamefont {V.}~\bibnamefont {Zwiller}},\ }\bibfield  {title} {\bibinfo
  {title} {Superconducting nanowire single-photon detectors: A perspective on
  evolution, state-of-the-art, future developments, and applications},\ }\href
  {https://doi.org/10.1063/5.0045990} {\bibfield  {journal} {\bibinfo
  {journal} {Applied Physics Letters}\ }\textbf {\bibinfo {volume} {118}},\
  \bibinfo {pages} {190502} (\bibinfo {year} {2021})}\BibitemShut {NoStop}%
\bibitem [{\citenamefont {Lamoreaux}\ \emph {et~al.}(2013)\citenamefont
  {Lamoreaux}, \citenamefont {van Bibber}, \citenamefont {Lehnert},\ and\
  \citenamefont {Carosi}}]{Lamoreaux_2013}%
  \BibitemOpen
  \bibfield  {author} {\bibinfo {author} {\bibfnamefont {S.~K.}\ \bibnamefont
  {Lamoreaux}}, \bibinfo {author} {\bibfnamefont {K.~A.}\ \bibnamefont {van
  Bibber}}, \bibinfo {author} {\bibfnamefont {K.~W.}\ \bibnamefont {Lehnert}},\
  and\ \bibinfo {author} {\bibfnamefont {G.}~\bibnamefont {Carosi}},\
  }\bibfield  {title} {\bibinfo {title} {Analysis of single-photon and linear
  amplifier detectors for microwave cavity dark matter axion searches},\ }\href
  {https://doi.org/10.1103/PhysRevD.88.035020} {\bibfield  {journal} {\bibinfo
  {journal} {Phys. Rev. D}\ }\textbf {\bibinfo {volume} {88}},\ \bibinfo
  {pages} {035020} (\bibinfo {year} {2013})}\BibitemShut {NoStop}%
\bibitem [{\citenamefont {McCuller}(2022)}]{mcculler2022singlephoton}%
  \BibitemOpen
  \bibfield  {author} {\bibinfo {author} {\bibfnamefont {L.}~\bibnamefont
  {McCuller}},\ }\href@noop {} {\bibinfo {title} {Single-photon signal sideband
  detection for high-power michelson interferometers}} (\bibinfo {year}
  {2022}),\ \Eprint {https://arxiv.org/abs/2211.04016} {arXiv:2211.04016
  [physics.ins-det]} \BibitemShut {NoStop}%
\bibitem [{\citenamefont {Marsili}\ \emph {et~al.}(2013)\citenamefont
  {Marsili}, \citenamefont {Verma}, \citenamefont {Stern}, \citenamefont
  {Harrington}, \citenamefont {Lita}, \citenamefont {Gerrits}, \citenamefont
  {Vayshenker}, \citenamefont {Baek}, \citenamefont {Shaw}, \citenamefont
  {Mirin},\ and\ \citenamefont {Nam}}]{Marsili_2013}%
  \BibitemOpen
  \bibfield  {author} {\bibinfo {author} {\bibfnamefont {F.}~\bibnamefont
  {Marsili}}, \bibinfo {author} {\bibfnamefont {V.~B.}\ \bibnamefont {Verma}},
  \bibinfo {author} {\bibfnamefont {J.~A.}\ \bibnamefont {Stern}}, \bibinfo
  {author} {\bibfnamefont {S.}~\bibnamefont {Harrington}}, \bibinfo {author}
  {\bibfnamefont {A.~E.}\ \bibnamefont {Lita}}, \bibinfo {author}
  {\bibfnamefont {T.}~\bibnamefont {Gerrits}}, \bibinfo {author} {\bibfnamefont
  {I.}~\bibnamefont {Vayshenker}}, \bibinfo {author} {\bibfnamefont
  {B.}~\bibnamefont {Baek}}, \bibinfo {author} {\bibfnamefont {M.~D.}\
  \bibnamefont {Shaw}}, \bibinfo {author} {\bibfnamefont {R.~P.}\ \bibnamefont
  {Mirin}},\ and\ \bibinfo {author} {\bibfnamefont {S.~W.}\ \bibnamefont
  {Nam}},\ }\bibfield  {title} {\bibinfo {title} {Detecting single infrared
  photons with 93{\%} system efficiency},\ }\href
  {https://doi.org/10.1038/nphoton.2013.13} {\bibfield  {journal} {\bibinfo
  {journal} {Nature Photonics}\ }\textbf {\bibinfo {volume} {7}},\ \bibinfo
  {pages} {210} (\bibinfo {year} {2013})}\BibitemShut {NoStop}%
\bibitem [{\citenamefont {Reddy}\ \emph {et~al.}(2020)\citenamefont {Reddy},
  \citenamefont {Nerem}, \citenamefont {Nam}, \citenamefont {Mirin},\ and\
  \citenamefont {Verma}}]{Reddy_20}%
  \BibitemOpen
  \bibfield  {author} {\bibinfo {author} {\bibfnamefont {D.~V.}\ \bibnamefont
  {Reddy}}, \bibinfo {author} {\bibfnamefont {R.~R.}\ \bibnamefont {Nerem}},
  \bibinfo {author} {\bibfnamefont {S.~W.}\ \bibnamefont {Nam}}, \bibinfo
  {author} {\bibfnamefont {R.~P.}\ \bibnamefont {Mirin}},\ and\ \bibinfo
  {author} {\bibfnamefont {V.~B.}\ \bibnamefont {Verma}},\ }\bibfield  {title}
  {\bibinfo {title} {Superconducting nanowire single-photon detectors with 98\%
  system detection efficiency at 1550 nm},\ }\href
  {https://doi.org/10.1364/OPTICA.400751} {\bibfield  {journal} {\bibinfo
  {journal} {Optica}\ }\textbf {\bibinfo {volume} {7}},\ \bibinfo {pages}
  {1649} (\bibinfo {year} {2020})}\BibitemShut {NoStop}%
\bibitem [{\citenamefont {Verma}\ \emph {et~al.}(2021)\citenamefont {Verma},
  \citenamefont {Korzh}, \citenamefont {Walter}, \citenamefont {Lita},
  \citenamefont {Briggs}, \citenamefont {Colangelo}, \citenamefont {Zhai},
  \citenamefont {Wollman}, \citenamefont {Beyer}, \citenamefont {Allmaras},
  \citenamefont {Vora}, \citenamefont {Zhu}, \citenamefont {Schmidt},
  \citenamefont {Kozorezov}, \citenamefont {Berggren}, \citenamefont {Mirin},
  \citenamefont {Nam},\ and\ \citenamefont {Shaw}}]{Verman_2021}%
  \BibitemOpen
  \bibfield  {author} {\bibinfo {author} {\bibfnamefont {V.}~\bibnamefont
  {Verma}}, \bibinfo {author} {\bibfnamefont {B.}~\bibnamefont {Korzh}},
  \bibinfo {author} {\bibfnamefont {A.}~\bibnamefont {Walter}}, \bibinfo
  {author} {\bibfnamefont {A.}~\bibnamefont {Lita}}, \bibinfo {author}
  {\bibfnamefont {R.}~\bibnamefont {Briggs}}, \bibinfo {author} {\bibfnamefont
  {M.}~\bibnamefont {Colangelo}}, \bibinfo {author} {\bibfnamefont
  {Y.}~\bibnamefont {Zhai}}, \bibinfo {author} {\bibfnamefont {E.}~\bibnamefont
  {Wollman}}, \bibinfo {author} {\bibfnamefont {A.}~\bibnamefont {Beyer}},
  \bibinfo {author} {\bibfnamefont {J.}~\bibnamefont {Allmaras}}, \bibinfo
  {author} {\bibfnamefont {H.}~\bibnamefont {Vora}}, \bibinfo {author}
  {\bibfnamefont {D.}~\bibnamefont {Zhu}}, \bibinfo {author} {\bibfnamefont
  {E.}~\bibnamefont {Schmidt}}, \bibinfo {author} {\bibfnamefont
  {A.}~\bibnamefont {Kozorezov}}, \bibinfo {author} {\bibfnamefont
  {K.}~\bibnamefont {Berggren}}, \bibinfo {author} {\bibfnamefont
  {R.}~\bibnamefont {Mirin}}, \bibinfo {author} {\bibfnamefont
  {S.}~\bibnamefont {Nam}},\ and\ \bibinfo {author} {\bibfnamefont
  {M.}~\bibnamefont {Shaw}},\ }\bibfield  {title} {\bibinfo {title}
  {{Single-photon detection in the mid-infrared up to 10 um wavelength using
  tungsten silicide superconducting nanowire detectors}},\ }\href
  {https://doi.org/10.1063/5.0048049} {\bibfield  {journal} {\bibinfo
  {journal} {APL Photonics}\ }\textbf {\bibinfo {volume} {6}},\ \bibinfo
  {pages} {056101} (\bibinfo {year} {2021})}\BibitemShut {NoStop}%
\bibitem [{\citenamefont {Grote}\ and\ \citenamefont {(the LIGO
  Scientific~Collaboration)}(2010)}]{Grote_2010}%
  \BibitemOpen
  \bibfield  {author} {\bibinfo {author} {\bibfnamefont {H.}~\bibnamefont
  {Grote}}\ and\ \bibinfo {author} {\bibnamefont {(the LIGO
  Scientific~Collaboration)}},\ }\bibfield  {title} {\bibinfo {title} {The
  {GEO} 600 status},\ }\href {https://doi.org/10.1088/0264-9381/27/8/084003}
  {\bibfield  {journal} {\bibinfo  {journal} {Classical and Quantum Gravity}\
  }\textbf {\bibinfo {volume} {27}},\ \bibinfo {pages} {084003} (\bibinfo
  {year} {2010})}\BibitemShut {NoStop}%
\bibitem [{\citenamefont {Abbott}\ \emph {et~al.}(2009)\citenamefont {Abbott}
  \emph {et~al.}}]{LIGO_2009}%
  \BibitemOpen
  \bibfield  {author} {\bibinfo {author} {\bibfnamefont {B.~P.}\ \bibnamefont
  {Abbott}} \emph {et~al.},\ }\bibfield  {title} {\bibinfo {title} {{LIGO: the
  Laser Interferometer Gravitational-Wave Observatory}},\ }\href
  {https://doi.org/10.1088/0034-4885/72/7/076901} {\bibfield  {journal}
  {\bibinfo  {journal} {Reports on Progress in Physics}\ }\textbf {\bibinfo
  {volume} {72}},\ \bibinfo {pages} {076901} (\bibinfo {year}
  {2009})}\BibitemShut {NoStop}%
\bibitem [{\citenamefont {Abbott}\ \emph {et~al.}(2016)\citenamefont {Abbott}
  \emph {et~al.}}]{O1instrPRL2016}%
  \BibitemOpen
  \bibfield  {author} {\bibinfo {author} {\bibfnamefont {B.~P.}\ \bibnamefont
  {Abbott}} \emph {et~al.} (\bibinfo {collaboration} {LIGO Scientific
  Collaboration and Virgo Collaboration}),\ }\bibfield  {title} {\bibinfo
  {title} {{GW150914: The Advanced LIGO Detectors in the Era of First
  Discoveries}},\ }\href {https://doi.org/10.1103/PhysRevLett.116.131103}
  {\bibfield  {journal} {\bibinfo  {journal} {Phys. Rev. Lett.}\ }\textbf
  {\bibinfo {volume} {116}},\ \bibinfo {pages} {131103} (\bibinfo {year}
  {2016})}\BibitemShut {NoStop}%
\bibitem [{\citenamefont {Martynov}\ and\ \citenamefont
  {Miao}(2020)}]{Martynov_2020}%
  \BibitemOpen
  \bibfield  {author} {\bibinfo {author} {\bibfnamefont {D.}~\bibnamefont
  {Martynov}}\ and\ \bibinfo {author} {\bibfnamefont {H.}~\bibnamefont
  {Miao}},\ }\bibfield  {title} {\bibinfo {title} {Quantum-enhanced
  interferometry for axion searches},\ }\href
  {https://doi.org/10.1103/PhysRevD.101.095034} {\bibfield  {journal} {\bibinfo
   {journal} {Phys. Rev. D}\ }\textbf {\bibinfo {volume} {101}},\ \bibinfo
  {pages} {095034} (\bibinfo {year} {2020})}\BibitemShut {NoStop}%
\bibitem [{\citenamefont {Staley}\ \emph {et~al.}(2014)\citenamefont {Staley},
  \citenamefont {Martynov}, \citenamefont {Abbott}, \citenamefont {Adhikari},
  \citenamefont {Arai}, \citenamefont {Ballmer}, \citenamefont {Barsotti},
  \citenamefont {Brooks}, \citenamefont {DeRosa}, \citenamefont {Dwyer},
  \citenamefont {Effler}, \citenamefont {Evans}, \citenamefont {Fritschel},
  \citenamefont {Frolov}, \citenamefont {Gray}, \citenamefont {Guido},
  \citenamefont {Gustafson}, \citenamefont {Heintze}, \citenamefont {Hoak},
  \citenamefont {Izumi}, \citenamefont {Kawabe}, \citenamefont {King},
  \citenamefont {Kissel}, \citenamefont {Kokeyama}, \citenamefont {Landry},
  \citenamefont {McClelland}, \citenamefont {Miller}, \citenamefont {Mullavey},
  \citenamefont {O'Reilly}, \citenamefont {Rollins}, \citenamefont {Sanders},
  \citenamefont {Schofield}, \citenamefont {Sigg}, \citenamefont {Slagmolen},
  \citenamefont {Smith-Lefebvre}, \citenamefont {Vajente}, \citenamefont
  {Ward},\ and\ \citenamefont {Wipf}}]{Staley_2014}%
  \BibitemOpen
  \bibfield  {author} {\bibinfo {author} {\bibfnamefont {A.}~\bibnamefont
  {Staley}}, \bibinfo {author} {\bibfnamefont {D.}~\bibnamefont {Martynov}},
  \bibinfo {author} {\bibfnamefont {R.}~\bibnamefont {Abbott}}, \bibinfo
  {author} {\bibfnamefont {R.~X.}\ \bibnamefont {Adhikari}}, \bibinfo {author}
  {\bibfnamefont {K.}~\bibnamefont {Arai}}, \bibinfo {author} {\bibfnamefont
  {S.}~\bibnamefont {Ballmer}}, \bibinfo {author} {\bibfnamefont
  {L.}~\bibnamefont {Barsotti}}, \bibinfo {author} {\bibfnamefont {A.~F.}\
  \bibnamefont {Brooks}}, \bibinfo {author} {\bibfnamefont {R.~T.}\
  \bibnamefont {DeRosa}}, \bibinfo {author} {\bibfnamefont {S.}~\bibnamefont
  {Dwyer}}, \bibinfo {author} {\bibfnamefont {A.}~\bibnamefont {Effler}},
  \bibinfo {author} {\bibfnamefont {M.}~\bibnamefont {Evans}}, \bibinfo
  {author} {\bibfnamefont {P.}~\bibnamefont {Fritschel}}, \bibinfo {author}
  {\bibfnamefont {V.~V.}\ \bibnamefont {Frolov}}, \bibinfo {author}
  {\bibfnamefont {C.}~\bibnamefont {Gray}}, \bibinfo {author} {\bibfnamefont
  {C.~J.}\ \bibnamefont {Guido}}, \bibinfo {author} {\bibfnamefont
  {R.}~\bibnamefont {Gustafson}}, \bibinfo {author} {\bibfnamefont
  {M.}~\bibnamefont {Heintze}}, \bibinfo {author} {\bibfnamefont
  {D.}~\bibnamefont {Hoak}}, \bibinfo {author} {\bibfnamefont {K.}~\bibnamefont
  {Izumi}}, \bibinfo {author} {\bibfnamefont {K.}~\bibnamefont {Kawabe}},
  \bibinfo {author} {\bibfnamefont {E.~J.}\ \bibnamefont {King}}, \bibinfo
  {author} {\bibfnamefont {J.~S.}\ \bibnamefont {Kissel}}, \bibinfo {author}
  {\bibfnamefont {K.}~\bibnamefont {Kokeyama}}, \bibinfo {author}
  {\bibfnamefont {M.}~\bibnamefont {Landry}}, \bibinfo {author} {\bibfnamefont
  {D.~E.}\ \bibnamefont {McClelland}}, \bibinfo {author} {\bibfnamefont
  {J.}~\bibnamefont {Miller}}, \bibinfo {author} {\bibfnamefont
  {A.}~\bibnamefont {Mullavey}}, \bibinfo {author} {\bibfnamefont
  {B.}~\bibnamefont {O'Reilly}}, \bibinfo {author} {\bibfnamefont {J.~G.}\
  \bibnamefont {Rollins}}, \bibinfo {author} {\bibfnamefont {J.~R.}\
  \bibnamefont {Sanders}}, \bibinfo {author} {\bibfnamefont {R.~M.~S.}\
  \bibnamefont {Schofield}}, \bibinfo {author} {\bibfnamefont {D.}~\bibnamefont
  {Sigg}}, \bibinfo {author} {\bibfnamefont {B.~J.~J.}\ \bibnamefont
  {Slagmolen}}, \bibinfo {author} {\bibfnamefont {N.~D.}\ \bibnamefont
  {Smith-Lefebvre}}, \bibinfo {author} {\bibfnamefont {G.}~\bibnamefont
  {Vajente}}, \bibinfo {author} {\bibfnamefont {R.~L.}\ \bibnamefont {Ward}},\
  and\ \bibinfo {author} {\bibfnamefont {C.}~\bibnamefont {Wipf}},\ }\bibfield
  {title} {\bibinfo {title} {{Achieving resonance in the Advanced LIGO
  gravitational-wave interferometer}},\ }\href
  {https://doi.org/10.1088/0264-9381/31/24/245010} {\bibfield  {journal}
  {\bibinfo  {journal} {Classical and Quantum Gravity}\ }\textbf {\bibinfo
  {volume} {31}},\ \bibinfo {pages} {245010} (\bibinfo {year}
  {2014})}\BibitemShut {NoStop}%
\bibitem [{\citenamefont {Izumi}\ \emph {et~al.}(2012)\citenamefont {Izumi},
  \citenamefont {Arai}, \citenamefont {Barr}, \citenamefont {Betzwieser},
  \citenamefont {Brooks}, \citenamefont {Dahl}, \citenamefont {Doravari},
  \citenamefont {Driggers}, \citenamefont {Korth}, \citenamefont {Miao},
  \citenamefont {Rollins}, \citenamefont {Vass}, \citenamefont
  {Yeaton-Massey},\ and\ \citenamefont {Adhikari}}]{Izumi_12}%
  \BibitemOpen
  \bibfield  {author} {\bibinfo {author} {\bibfnamefont {K.}~\bibnamefont
  {Izumi}}, \bibinfo {author} {\bibfnamefont {K.}~\bibnamefont {Arai}},
  \bibinfo {author} {\bibfnamefont {B.}~\bibnamefont {Barr}}, \bibinfo {author}
  {\bibfnamefont {J.}~\bibnamefont {Betzwieser}}, \bibinfo {author}
  {\bibfnamefont {A.}~\bibnamefont {Brooks}}, \bibinfo {author} {\bibfnamefont
  {K.}~\bibnamefont {Dahl}}, \bibinfo {author} {\bibfnamefont {S.}~\bibnamefont
  {Doravari}}, \bibinfo {author} {\bibfnamefont {J.~C.}\ \bibnamefont
  {Driggers}}, \bibinfo {author} {\bibfnamefont {W.~Z.}\ \bibnamefont {Korth}},
  \bibinfo {author} {\bibfnamefont {H.}~\bibnamefont {Miao}}, \bibinfo {author}
  {\bibfnamefont {J.}~\bibnamefont {Rollins}}, \bibinfo {author} {\bibfnamefont
  {S.}~\bibnamefont {Vass}}, \bibinfo {author} {\bibfnamefont {D.}~\bibnamefont
  {Yeaton-Massey}},\ and\ \bibinfo {author} {\bibfnamefont {R.~X.}\
  \bibnamefont {Adhikari}},\ }\bibfield  {title} {\bibinfo {title} {Multicolor
  cavity metrology},\ }\href {https://doi.org/10.1364/JOSAA.29.002092}
  {\bibfield  {journal} {\bibinfo  {journal} {J. Opt. Soc. Am. A}\ }\textbf
  {\bibinfo {volume} {29}},\ \bibinfo {pages} {2092} (\bibinfo {year}
  {2012})}\BibitemShut {NoStop}%
\bibitem [{\citenamefont {Mullavey}\ \emph {et~al.}(2012)\citenamefont
  {Mullavey}, \citenamefont {Slagmolen}, \citenamefont {Miller}, \citenamefont
  {Evans}, \citenamefont {Fritschel}, \citenamefont {Sigg}, \citenamefont
  {Waldman}, \citenamefont {Shaddock},\ and\ \citenamefont
  {McClelland}}]{Mullavey:12}%
  \BibitemOpen
  \bibfield  {author} {\bibinfo {author} {\bibfnamefont {A.~J.}\ \bibnamefont
  {Mullavey}}, \bibinfo {author} {\bibfnamefont {B.~J.~J.}\ \bibnamefont
  {Slagmolen}}, \bibinfo {author} {\bibfnamefont {J.}~\bibnamefont {Miller}},
  \bibinfo {author} {\bibfnamefont {M.}~\bibnamefont {Evans}}, \bibinfo
  {author} {\bibfnamefont {P.}~\bibnamefont {Fritschel}}, \bibinfo {author}
  {\bibfnamefont {D.}~\bibnamefont {Sigg}}, \bibinfo {author} {\bibfnamefont
  {S.~J.}\ \bibnamefont {Waldman}}, \bibinfo {author} {\bibfnamefont {D.~A.}\
  \bibnamefont {Shaddock}},\ and\ \bibinfo {author} {\bibfnamefont {D.~E.}\
  \bibnamefont {McClelland}},\ }\bibfield  {title} {\bibinfo {title}
  {Arm-length stabilisation for interferometric gravitational-wave detectors
  using frequency-doubled auxiliary lasers},\ }\href
  {https://doi.org/10.1364/OE.20.000081} {\bibfield  {journal} {\bibinfo
  {journal} {Opt. Express}\ }\textbf {\bibinfo {volume} {20}},\ \bibinfo
  {pages} {81} (\bibinfo {year} {2012})}\BibitemShut {NoStop}%
\bibitem [{\citenamefont {Budker}\ \emph {et~al.}(2014)\citenamefont {Budker},
  \citenamefont {Graham}, \citenamefont {Ledbetter}, \citenamefont
  {Rajendran},\ and\ \citenamefont {Sushkov}}]{Budker_Casper_2014}%
  \BibitemOpen
  \bibfield  {author} {\bibinfo {author} {\bibfnamefont {D.}~\bibnamefont
  {Budker}}, \bibinfo {author} {\bibfnamefont {P.~W.}\ \bibnamefont {Graham}},
  \bibinfo {author} {\bibfnamefont {M.}~\bibnamefont {Ledbetter}}, \bibinfo
  {author} {\bibfnamefont {S.}~\bibnamefont {Rajendran}},\ and\ \bibinfo
  {author} {\bibfnamefont {A.~O.}\ \bibnamefont {Sushkov}},\ }\bibfield
  {title} {\bibinfo {title} {Proposal for a cosmic axion spin precession
  experiment (casper)},\ }\href {https://doi.org/10.1103/PhysRevX.4.021030}
  {\bibfield  {journal} {\bibinfo  {journal} {Phys. Rev. X}\ }\textbf {\bibinfo
  {volume} {4}},\ \bibinfo {pages} {021030} (\bibinfo {year}
  {2014})}\BibitemShut {NoStop}%
\bibitem [{\citenamefont {Schumaker}\ and\ \citenamefont
  {Caves}(1985)}]{Caves_1985}%
  \BibitemOpen
  \bibfield  {author} {\bibinfo {author} {\bibfnamefont {B.~L.}\ \bibnamefont
  {Schumaker}}\ and\ \bibinfo {author} {\bibfnamefont {C.~M.}\ \bibnamefont
  {Caves}},\ }\bibfield  {title} {\bibinfo {title} {New formalism for
  two-photon quantum optics. ii. mathematical foundation and compact
  notation},\ }\href {https://doi.org/10.1103/PhysRevA.31.3093} {\bibfield
  {journal} {\bibinfo  {journal} {Phys. Rev. A}\ }\textbf {\bibinfo {volume}
  {31}},\ \bibinfo {pages} {3093} (\bibinfo {year} {1985})}\BibitemShut
  {NoStop}%
\bibitem [{\citenamefont {Schnabel}(2017)}]{Schnabel_2017}%
  \BibitemOpen
  \bibfield  {author} {\bibinfo {author} {\bibfnamefont {R.}~\bibnamefont
  {Schnabel}},\ }\bibfield  {title} {\bibinfo {title} {Squeezed states of light
  and their applications in laser interferometers},\ }\href
  {https://doi.org/10.1016/j.physrep.2017.04.001} {\bibfield  {journal}
  {\bibinfo  {journal} {Physics Reports}\ }\textbf {\bibinfo {volume} {684}},\
  \bibinfo {pages} {1} (\bibinfo {year} {2017})}\BibitemShut {NoStop}%
\bibitem [{\citenamefont {Dixit}\ \emph {et~al.}(2021)\citenamefont {Dixit},
  \citenamefont {Chakram}, \citenamefont {He}, \citenamefont {Agrawal},
  \citenamefont {Naik}, \citenamefont {Schuster},\ and\ \citenamefont
  {Chou}}]{Dixit_2021}%
  \BibitemOpen
  \bibfield  {author} {\bibinfo {author} {\bibfnamefont {A.~V.}\ \bibnamefont
  {Dixit}}, \bibinfo {author} {\bibfnamefont {S.}~\bibnamefont {Chakram}},
  \bibinfo {author} {\bibfnamefont {K.}~\bibnamefont {He}}, \bibinfo {author}
  {\bibfnamefont {A.}~\bibnamefont {Agrawal}}, \bibinfo {author} {\bibfnamefont
  {R.~K.}\ \bibnamefont {Naik}}, \bibinfo {author} {\bibfnamefont {D.~I.}\
  \bibnamefont {Schuster}},\ and\ \bibinfo {author} {\bibfnamefont
  {A.}~\bibnamefont {Chou}},\ }\bibfield  {title} {\bibinfo {title} {Searching
  for dark matter with a superconducting qubit},\ }\href
  {https://doi.org/10.1103/PhysRevLett.126.141302} {\bibfield  {journal}
  {\bibinfo  {journal} {Phys. Rev. Lett.}\ }\textbf {\bibinfo {volume} {126}},\
  \bibinfo {pages} {141302} (\bibinfo {year} {2021})}\BibitemShut {NoStop}%
\end{thebibliography}%

\end{document}